# Approaching a Minimal Topological Electronic Structure in Antiferromagnetic Topological Insulator MnBi$_2$Te$_4$ via Surface Modification

Aiji Liang[1,2,*], Cheng Chen[3,*], Huijun Zheng[1], Wei Xia[1,2], Kui Huang[1], Liyang Wei[1], Haifeng Yang[1], Yujie Chen[4], Xin Zhang[1], Xuguang Xu[1,2], Meixiao Wang[1,2], Yanfeng Guo[1,2], Lexian Yang[4,5], Zhongkai Liu[1,2,‡], Yulin Chen[1,2,3,4,‡]

[1]*School of Physical Science and Technology, ShanghaiTech University, Shanghai 201210, China*

[2]*ShanghaiTech Laboratory for Topological Physics, Shanghai 200031, P. R. China*

[3]*Clarendon Laboratory, Department of Physics, University of Oxford, Oxford OX1 3PU, U.K.*

[4]*State Key Laboratory of Low Dimensional Quantum Physics, Department of Physics, Tsinghua University, Beijing 100084, China*

[5]*Frontier Science Center for Quantum Information, Beijing 100084, China*

**Abstract:**

The topological electronic structure plays a central role in the non-trivial physical properties in topological quantum materials. A minimal, "hydrogen-atom-like" topological electronic structure is desired for researches. In this work, we demonstrate an effort towards the realization of such a system in the intrinsic magnetic topological insulator MnBi$_2$Te$_4$, by manipulating the topological surface state (TSS) via surface modification. Using high resolution laser- and synchrotron-based angle-resolved photoemission spectroscopy (ARPES), we found the TSS in MnBi$_2$Te$_4$ is heavily hybridized with a trivial Rashba-type surface state (RSS), which could be efficiently removed by the in *situ* surface potassium (K) dosing. By employing multiple experimental methods to characterize K dosed surface, we attribute such **a modification to the electrochemical reactions of K clusters on the surface. Our work not only gives a clear band assignment in MnBi$_2$Te$_4$, but also provides possible new routes in accentuating the topological behavior in the magnetic topological quantum materials.**

In the past decade, the investigation of topological quantum material (TQM) has become the frontier of condensed matter physics as well as material science [1-3]. TQM provides us a unique playground to investigate fascinating topological physics as well as platform to achieve important device applications. In the search of ideal TQMs with desired performance (e.g., high temperature quantum anomalous Hall effect (QAHE)), the optimization of the topological electronic structure is essential. A minimal topological electronic structure (i.e., the simplest topological electronic structure dominates the density of states near the Fermi level ($E_F$), e.g., topological insulator with only a single Dirac cone surface state or topological Weyl semimetal with only minimal number of Weyl points at/near $E_F$) is desired for the research due to its simplicity, the accentuated topological behaviors and thus feasible applications. To realize such a system, one can either search for the new materials with ideal electronic structure from the database [4-8] or "band-engineering" the existing materials through chemical doping, straining, gating, etc. [9].

Among the TQMs, the magnetic topological insulator (MTI) are important research platforms in realizing QAHE [10-11, 19-20], axion insulator state (AIS) [21], controllable phase transition between AIS and Chern insulator state [21] and high Chern number Chern insulator state [22]. To date, massive researches show [12-54] that $MnBi_2Te_4$ is the most representative intrinsic MTI due to its easy synthesis, relatively high antiferromagnetic (AFM) ordering temperature ($T_N$~25K [13-15, 23-24]) and moderate bulk band gap (~0.2 eV, [25-29]). The theoretical calculations predict a single topological Dirac cone surface state at Brillouin zone center (Γ), close to realizing a minimal topological electronic structure [25-29]. However, despite the great advances in transports [19-22], there are still conflicting spectroscopy experiments that need to be addressed, including the assignment of bulk and surface states near $E_F$, the exchange gap size of TSS (~60 meV from calculation [18, 25-26, 32], diminished gap from laser-based ARPES [31-33] and tens of meV from synchrotron-based ARPES [34-36]). In addition, the $T_c$ of the realized QAHE (~1.5 K) in $MnBi_2Te_4$ is relatively low comparing to $T_N$ [19-20]. Therefore, a further optimization of the electronic structure to achieve a minimal MTI electronic structure would suppress undesired conducting channels and may help to increase $T_c$.

In this work, we show that the solution of band assignment and an effort towards the realization of a minimal MTI electronic structure can be coincidently obtained by in *situ* potassium atom dosing (K dosing) in $MnBi_2Te_4$, which is unveiled by our laser- and synchrotron- based angle-resolved photoemission spectroscopy (ARPES) studies. We found that the TSS in $MnBi_2Te_4$ is heavily hybridized with a Rashba-type surface state (RSS, the schematic of the unhybridized bandstructure is shown in Fig. 1a while the hybridized bandstructure is shown in Fig. 1b), which could be efficiently manipulated through a two-stage cross-over process of continuous K dosing: in stage 1, the hybridization of TSS and RSS is modified and the hybridization gap is eventually closed; while in stage 2, the RSS fades away, approaching a minimal MTI electronic structure with only TSS and inverted bulk bands (Fig. 1c) near $E_F$. Combining core level analysis, scanning tunneling microscopy (STM) and low energy electron diffraction (LEED) results, we attribute such two-stage dosing-evolution to the adsorption of K atoms and electrochemical reaction of K clusters on the $MnBi_2Te_4$ surface, respectively. Our work clarifies the assignment of the bulk and surface states around the $E_F$ in $MnBi_2Te_4$. Moreover, the simple manipulation and the effort towards realizing a minimal MTI electronic structure in $MnBi_2Te_4$ via K dosing also provides new routes that may help to realize or improve novel topological phenomena in both transport experiments and device applications in MTI families.

We present the laser circular dichroism (CD) ARPES measurement of pristine $MnBi_2Te_4$ in Figure 1 (~80 K above $T_N$). Clear CD contrast is observed along both $\bar{\Gamma}$-$\bar{K}$ (Figs. 1d-g) and $\bar{\Gamma}$-$\bar{M}$ directions (SI1). The bulk and surface states can be therefore assigned accordingly as labelled in Fig. 1g. The SS1 is the Dirac cone-type TSS and SS2 is the outer part of a RSS, which hybridizes with SS1 and has been interpreted as part of TSS in previous reports [31-33]. The different origin of the two surface states is evidenced by the dramatic change of the CD contrast (Fig. 1g). The discontinuity of the CD intensity ('kink'-like feature) marks the hybridization between SS1 and SS2 (labelled as Hyb. Pos. in Figs. 1b, g). SS3 is the inner part of the RSS [38] which merges into the bulk conduction band (CB) after the hybridized gap opens (Figs. 1b, g).

The observed surface states and hybridization can be effectively modified by the surface K dosing (Figure 2). The measurement is performed at ~18K (below $T_N$) where the CB and the valence band (VB) split into CB'/CB'' and VB'/VB'' due to exchange coupling (Fig. 2b and Ref. 31). Figs. 2a-g show the systematic evolution of band dispersions along $\bar{\Gamma}$-$\bar{M}$ with sequential K dosing. The overall electronic structure undergoes a non-monotonic 'V' shape shift in energy (Fig. 2h), suggesting a two-stage effective doping (upward/downward shift of the Fermi level). The energy shift of the surface states (SS1, SS2, SS3) is much larger comparing to the bulk bands (CB'/CB'' and VB'/VB'') (Fig. 2h), indicating their surface and bulk origins. The two-stage transition is a crossover process where we can find a critical K dosing to separate them which is denoted by the removal of the hybridization gap between SS1, SS2 and SS3. This is clear when one traces SS3 in Fig. 2h and is more evident in the curvature plot [55] in Figs. 2e-f where SS3 crosses CB' and develops extra states in between CB' and SS1 (now SS1 and SS3 form the TSS). The observed hybridization gap closing thus supports our assignment of TSS in Figure 1 (see SI2 for more discussions).

In Figure 3, we demonstrate that exchange splitting of the bulk bands and TSS with diminished gap are preserved in $MnBi_2Te_4$ after heavy K dosage. This is unveiled by side-to-side comparison of the temperature dependent spectra between pristine (Figs. 3a-d) and heavily dosed samples (Figs. 3e-h). The band splitting closing temperature in both cases is ~28K [31], close to $T_N$. Therefore, the K atoms are only deposited on the surface since the intercalation of K atoms between the septuple layers will expand layer distance and lower the $T_N$, as observed in $(MnBi_2Te_4)_m(Bi_2Te_3)_n$ heterostructures [39-45]. Despite the slight changes in the bulk bands and surface states (the effective mass of CB' and velocity of linear SS1 near Dirac point (DP) change from 0.13 $m_e$, 2.57 eV·Å (Fig. 3a) to 0.12 $m_e$, 3.67 eV·Å (Fig. 3e)), heavily K dosed $MnBi_2Te_4$ preserves the key topological features of an MTI, i.e. inverted bulk bands which split below $T_N$, and single Dirac cone TSS at Γ point. Therefore, we approach a minimal topological electronic structure of an MTI via K dosage. In SI4 and 6, we show massive K dosing can bring CB bottom much closer to $E_F$, suggesting that we may achieve the minimal topological electronic structure of an MTI by back-gating, as shown in the schematic in SI5.

To understand the microscopic mechanism of such a K dosing evolution, we employ synchrotron-based ARPES to trace the core level changes during the process, as shown in Figure 4. Fig. 4a presents the evolution of Te 4d, Bi 5d and the K 3p core levels upon dosing. Two stages of evolution could be observed, consistent with the evolution of low energy band structure in Figure 2 (see SI4). The observed core level peaks first shift together towards higher binding energy (upward shift of the Fermi level) in stage 1, and shift backwards (downward shift of the Fermi level) in stage 2 together with the development of pronounced side peaks (such as the K 3p and Bi 5d core level peaks in Figs. 4a-b).

The microscopic mechanism of K deposition is a two-stage electrochemical reaction process. The different K coverage controls the reaction stage and leads to the two stages of K dosing. At stage 1, a small amount of K atoms is adsorbed on the as-cleaved $MnBi_2Te_4$ surface. The sparse surface K adatoms with negligible mutual interaction contribute to a single K 3p core level peak ($E_b$~19eV, Figs. 4a-b). The charge transfer from K atoms to $MnBi_2Te_4$ results in upward shift of the Fermi level and shifts the electronic bands and core level towards higher binding energy. Surface states (especially SS2 and SS3) are modified more significantly (Fig. 2a and SI4), suggesting their different origin comparing to the SS1. At stage 2, K atoms accumulate so that mutual interaction of K atoms is no longer negligible. The interaction between $K^+$ tends to drive them into segregated K cluster which increases the surface roughness and thus lowers the surface formation energy. The random K clusters could be observed by the STM topography (Fig. 4c). As a result, a second K 3p peak ($E_b$~18.5eV) originated from clustered K atoms (labelled as K 3p bulk, Figs. 4a-b) can be observed. The resulted clustering of K atoms reduces charge transfer from K atoms to $MnBi_2Te_4$ surface. Further, the large K amount allows a K-Te-Bi alloying as proposed in Fe [56], Ni [57] and Ti [58] dosed $Bi_2Se_3$, which may explain the simultaneous development of the less oxidized Bi 5d peak (labelled as Bi 5d') and Te 4d side peak (SI4). K dosing on the $Bi_2Te_3$ plate has also been investigated carefully, also showing similar 2-stage electrochemical reaction and stage 2 shows the reduction and alloying of Bi [59]. The reduced Bi containing alloy extracts electrons from surface of $MnBi_2Te_4$ so that free carriers decrease, leading to downward shift of the Fermi level (in comparison to the most electron doped state and even the pristine sample) and shifting the electronic bands and core level peaks towards $E_F$ (SI4). The existence of both K cluster and the K-Te-Bi alloy could be proved by heating the sample, which can effectively remove the K adatoms but not the clusters and alloys (SI6). We further rule out the possibility of surface reconstruction or K ordering [60-61] by measuring the LEED patterns before and after K dosage (Fig. 4d).

Based on the observations above, we summarize the low energy band structure near $E_F$ of $MnBi_2Te_4$ and its evolution upon K dosing in Figs. 4e-h. A topologically nontrivial TSS and a trivial RSS coexist (Fig. 4e) and hybridize with each other to give rise to the near $E_F$ band structure of pristine $MnBi_2Te_4$ (Fig. 4f). Upon K dosing (Figs. 4g-h), the adsorbed K atoms and subsequently formed K clusters and alloys drive upward shift of the Fermi level (Fig. 4g) and downward shift of the Fermi level (Fig. 4h) of $MnBi_2Te_4$. The randomly formed K clusters and alloys modifies surface potential greatly and finally suppresses the trivial surface state, leaving only TSS and inverted bulk bands, approaching the minimal topological electronic band structure of an MTI (Fig. 4h). With the DP further tuned closer to $E_F$ after massive K dosing (SI4 and 5), such a close-to-ideal MTI may host QAHE on the sample surface. We note that the similar RSS and evolution with surface K dosing has been observed in $MnBi_2Te_4$ films grown by molecular beam epitaxy (Xu, R. Z.; et al., unpublished experiments), suggesting the possibility to improve the QAHE in exfoliated films via surface modification (see further discussion in SI4-5).

The origin of RSS in MnBi$_2$Te$_4$ is different from the case in Bi$_2$Se$_3$ where extra modification by either vacuum exposure [62-63] or alkali atoms deposition are needed [62,64]. Instead, the RSS in MnBi$_2$Te$_4$ can be observed in fresh as-cleaved sample surface [32,37-38,54]. It can also be reproduced by calculations, though the energy position and shape have a considerable deviation from experiment [32,37]. After heavy K dosage, the K clusters and the Bi alloys bring great disordered potential at the MnBi$_2$Te$_4$ surface so as to break the translational symmetry. The lack of periodicity at the top surface would thus annihilate the RSS.

The simplified electronic structure and clear bands assignments of MnBi$_2$Te$_4$ would serve as a solid starting point to understand the key electronic structures such as the origin of TSS and its interplay with magnetism. Further, the modified MnBi$_2$Te$_4$ surface provides a "passivated" layer with both the almost minimal topological electronic structure as well as naturally formed electrodes (Bi-alloys) on top, such configuration allows future investigation of the quantum transport behavior as well as device application with intrinsic MTIs.

**Methods**

Single crystals of MnBi$_2$Te$_4$ were grown via a solid-state reaction method as described in earlier studies [14]. laser-based ARPES were measured at ShanghaiTech University with a DA30L analyser (Scienta Omicron) and 6.994 eV laser. The energy resolution was ~3.75 meV. Synchrotron-based ARPES data were collected in Beamline 7.0.2 and Beamline 10.0.1 of Advanced Light Source with R4000 analysers. The energy resolution was ~15 meV. The samples were cleaved *in situ* and measured under vacuum below $5 \times 10^{-11}$ Torr. Sequential in *situ* K dosing was employed in both measurements.

ASSOCIATED CONTENT

Supporting Information

The Supporting Information is available free of charge at http://pubs.acs.org.

SI1 Laser circular dichroism of MnBi$_2$Te$_4$

SI2 Volume of K dosing dependent band structure of MnBi$_2$Te$_4$ by laser-based ARPES

SI3 Temperature dependent conduction band splitting of pristine and massive K dosed MnBi$_2$Te$_4$

SI4 Volume of K dosing dependent band structure of MnBi$_2$Te$_4$ by synchrotron-based ARPES

SI5 Schematic of back gate on massive K dosed MnBi$_2$Te$_4$ film

SI6 Volume of K dosing dependent band structure of MnBi$_2$Te$_4$ after annealing

**Author information**

**Corresponding Authors**

Zhongkai Liu. E-mail: liuzhk@shanghaitech.edu.cn

Yulin Chen. E-mail: yulin.chen@physics.ox.ac.uk

**Author Contributions**

*Aiji Liang and Cheng Chen contributed equally in the work.

**Notes**

The authors declare no competing financial interest.

**Acknowledgments**

We thank Y. Han, Y. Xu, J. P. Liu and W. J. Shi for helpful discussions. We appreciate the help from E. Rotenberg, A. Bostwick, C. Jozwiak and S.-K. Mo for the measurements at Advanced Light Source. This research used resources of the Advanced Light Source, which is a DOE Office of Science User Facility under contract no. DE-AC02-05CH11231. Z. K. Liu acknowledges the National Key R&D program of China (Grant No. 2017YFA0305400). X. G. Xu acknowledges the Shanghai Technology Innovation Action Plan 2020-Integrated Circuit Technology Support Program (Project No. 20DZ1100605).

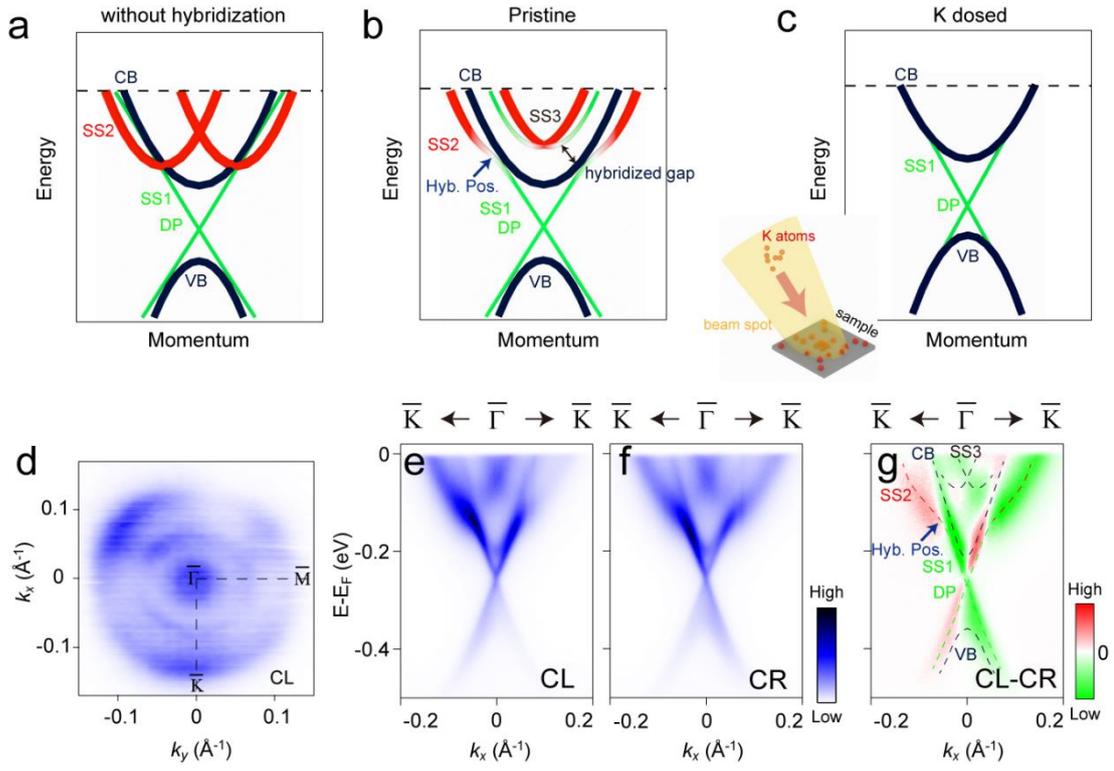

**Figure 1. Schematic of approaching a minimal topological electronic structure via K dosing and laser circular dichroism (CD) of MnBi$_2$Te$_4$.** (a) Schematic of band structure of MnBi$_2$Te$_4$ without the hybridization of Rashba-type surface state RSS (SS2) and topological surface state TSS (SS1). Conduction band (CB), valence band (VB) and Dirac point (DP) are marked. (b) Schematic of band structure of pristine MnBi$_2$Te$_4$, considering hybridization of SS1 and SS2. The hybridized gap opening rearranges surface states, generating Rashba-type SS3 [38]. Hybridization position (Hyb. Pos.) and hybridized gap are noted. (c) Schematic of band structure of massive K dosed MnBi$_2$Te$_4$. Inset shows the K dosing setup. (d) Fermi surface of MnBi$_2$Te$_4$ measured by circular left (CL) polarized photons along $\bar{\Gamma}$-$\bar{K}$. (e, f) Band dispersions along $\bar{\Gamma}$-$\bar{K}$ measured by CL and circular right (CR) polarized photons. (g) CD image obtained by the difference (CL-CR) of those in (e, f). The overlaid curves are guides for eyes. The measurement temperature is ~80 K.

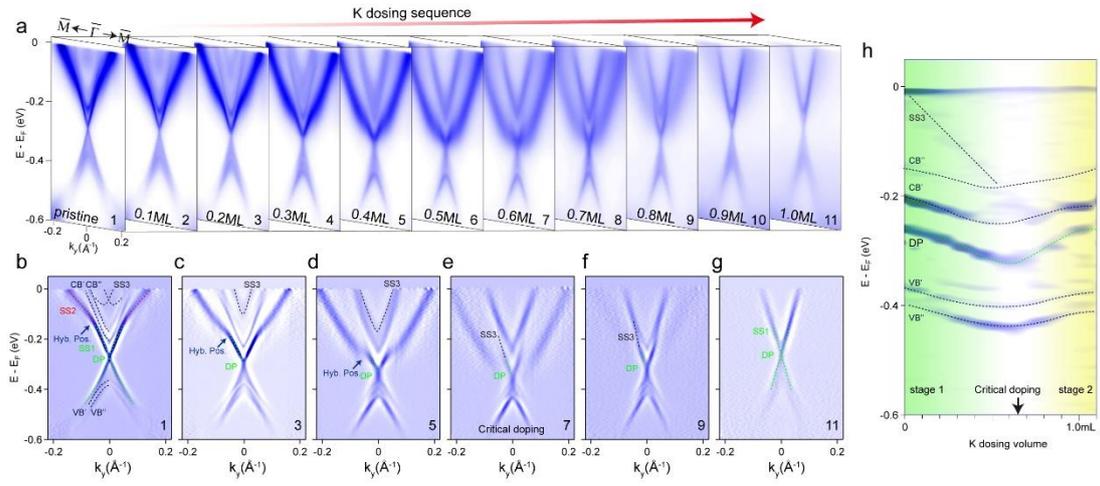

**Figure 2. Manipulating band structure of MnBi$_2$Te$_4$ by in *situ* K dosing**. (a) Stacked plots of band structure evolution along $\bar{\Gamma}$-$\bar{M}$ upon sequential K dosing. The dosage is continuous and the dosing condition remains the same for each round (ML, mono-layer. The K dosing volume estimation is discussed in SI4). The sample temperature is kept at 18 K (below T$_N$~25 K) for measurement and dosing. Noted the seemingly ultra-weak background of RSS in sequence 11 can be completely removed by further dosing. Please find SFigure 4 in SI4 for details. (b-g) Representative curvature plots from those in (a). Conduction (CB', CB'') and valence (VB', VB'') bands split due to the AFM ordering. (h) Curvature plot of the image formed by stacking the energy distributed curves (EDCs) at $k_y$=0 from those in (a). Bands, two-stage crossover dosing process and the critical doping are denoted.

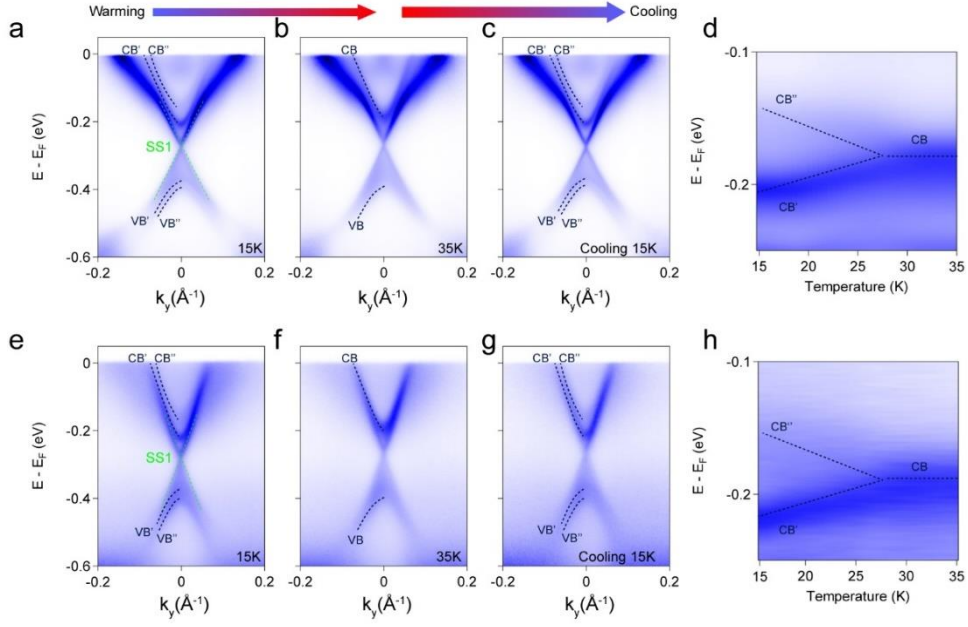

**Figure 3. Persistent bulk band splitting and diminished gap of TSS of MnBi$_2$Te$_4$ upon K dosing**. (a) Band dispersions along $\bar{\Gamma}$-$\bar{M}$ of pristine MnBi$_2$Te$_4$ measured at 15 K bellow T$_N$. Bulk band splitting due to AFM ordering are denoted accordingly. (b) same as (a) but measured at 35 K above T$_N$ (warming up from 15 K). (c) same as (a) but measured at 15 K (cooling down from 35 K). (d) Intensity map of the stacked of EDCs of the conduction band at $k_y$=0 at various temperatures, which is taken by warming up the sample from 15 K to 35 K. The closing of the conduction band splitting is ~28 K [31]. (e-h) same measurement as (a-d) but on the massive K dosed sample. See SI3 for the original EDCs in d and h.

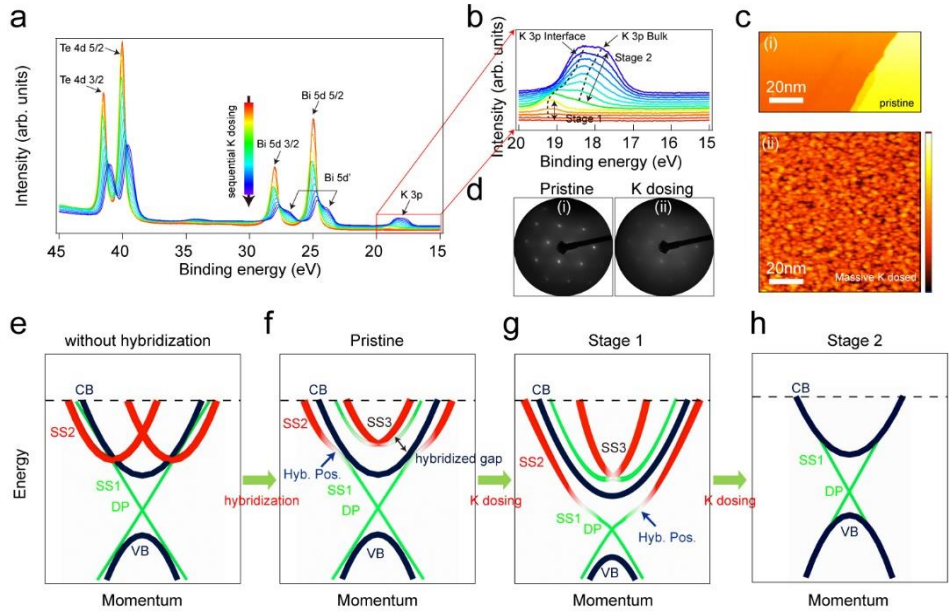

**Figure 4. Characterization of MnBi$_2$Te$_4$ upon K dosing and summary of band structure evolution.** (a) Evolution of Te 4d, Bi 5d and K 3p core levels upon K dosing. (b) Zoomed-in plot of the K 3p core level evolution. Interfacial and bulk K 3p core levels are marked. The two stages of K dosing are distinguished by the switch of shift direction of the interface K 3p core level together with the simultaneous appearance of bulk K 3p core level and less oxidized Bi 5d core levels (SI4). (c) Surface topography of pristine (i) and massive K dosed (ii) MnBi$_2$Te$_4$. K clusters in (ii) cover the whole surface. (d) LEED pattern taken from pristine (i) and massive K dosed (ii) MnBi$_2$Te$_4$ surface. (e) Schematic of band structure of MnBi$_2$Te$_4$ without the hybridization of RSS (SS2) and TSS (SS1). (f) Schematic of band structure of pristine MnBi$_2$Te$_4$, considering hybridization of SS1 and SS2. The hybridized gap opening rearranges surface states, generating Rashba-type SS3 [38]. (g) Schematic of band structure of MnBi$_2$Te$_4$ after a small volume of K dosing. Hybridization of SS1 and RSS is tuned. Surface chemical potential moves upwards (more electron doped). (h) Schematic of band structure of MnBi$_2$Te$_4$ after massive K dosing. RSS is suppressed. Surface chemical potential moves downwards (more hole doped). Noted that the CB bottom can be further

tuned towards Fermi level by massive K dosing. See SI4-6 for details.